# Multi-Weight Respecification of Scan-Specific Learning for Parallel Imaging

Hui Tao, Haifeng Wang, *Senior Member, IEEE*, Shanshan Wang, *Senior Member, IEEE*,
Dong Liang, *Senior Member, IEEE*, Xiaoling Xu, Qiegen Liu, *Senior Member, IEEE*

*Abstract*—Parallel imaging is widely used in magnetic resonance imaging as an acceleration technology. Traditional linear reconstruction methods in parallel imaging often suffer from noise amplification. Recently, a non-linear robust artificial-neural-network for k-space interpolation (RAKI) exhibits superior noise resilience over other linear methods. However, RAKI performs poorly at high acceleration rates, and needs a large amount of autocalibration signals as the training samples. In order to tackle these issues, we propose a multi-weight method that implements multiple weighting matrices on the undersampled data, named as MW-RAKI. Enforcing multiple weighted matrices on the measurements can effectively reduce the influence of noise and increase the data constraints. Furthermore, we incorporate the strategy of multiple weighting matrixes into a residual version of RAKI, and form MW-rRAKI. Experimental comparisons with the alternative methods demonstrated noticeably better reconstruction performances, particularly at high acceleration rates.

*Index Terms*—Parallel magnetic resonance imaging, imaging reconstruction, scan-specific learning, multi-weight respecification, k-space.

## I. INTRODUCTION

Magnetic resonance imaging (MRI) is an indispensable tool for disease diagnosis and treatment planning due to its noninvasive, radiation-free, and in-vivo imaging nature. One major shortcoming of MRI is the correspondingly slow data acquisition speed. To accelerate the acquisition, instead of acquiring the fully sampled k-space data, it can be subsampled at a frequency that is lower than the Nyquist rate [1]. Nevertheless, simple undersampling schemes result in aliasing artifacts in the reconstructed images, and many tissue structures in images are obscured by these artifacts [2].

Over the years, several methods have been proposed to accelerate MRI [3], [4]. One of the early approaches was the parallel imaging (PI), which exploits spatial sensitivity of multiple coils in conjunction with gradient encoding to reduce the data samples that are required for reconstruction [5]-[7]. Hence, the scanning time is greatly shortened. These methods can be divided into the image domain reconstruction and k-space domain reconstruction. Sensitivity encoding (SENSE) [8]-[10] is the typical algorithm of the image domain reconstruction. Given the prior knowledge of receiver sensitivity, it expands the overlapping part of the image with overlapping coils, and finally obtains a complete image through an inverse problem formulation. However, it suffers from increased noise amplification due to the ill-conditioning at higher acceleration rates [11]-[13]. K-space domain reconstruction methods include generalized autocalibrating partially parallel acquisitions (GRAPPA) [14], [15], simultaneous acquisition of spatial harmonics (SMASH) [16], [17], etc. For instance, the SPIRiT [18], [19] proposed by Lustig *et al.* [18] was fitted by linear weighted superposition of adjacent data, whether these data are collected or not. It is flexible and facilitates the use of regularization in the reconstruction process. LORAKS [20], [21] employs low-rank matrix to reconstruct undersampled images and accelerates MRI acquisitions through sophisticated priors on structure and redundancy in k-space data from multi-coil receive arrays. Unfortunately, these methods need iterative processing that requires large amounts of calculation.

Inspired by the success of deep learning, many researchers have investigated it for MR reconstruction in recent years [22]-[26]. Compared with traditional methods, MRI reconstruction algorithms based on deep learning are capable to recover high-quality images from undersampled acquisitions. At present, most of the reconstruction methods can be classified into three groups: End-to-end supervised deep learning [27]-[29], self-supervised deep learning [30], [31] and unsupervised deep learning [32], [33]. Supervised learning usually employs pairs of data samples but unsupervised learning systems are used when paired data labels are not available. Compared with supervised learning, unsupervised learning has better flexibility. Although supervised learning and unsupervised learning differ in some ways, they mainly learn the regularization from the undersampled k-space data to eliminate artifacts. However, these methods require large amounts of training data, containing fully-sampled data.

To address the issue that training data is insufficient, some self-supervised learning approaches such as scan-specific learning have been developed. For instance, Akçakaya *et al.* [34] designed robust artificial-neural-networks for k-space interpolation (RAKI) by using scan-specific deep learning [34]-[36]. Instead of training linear convolutional kernels from the autocalibration signal (*ACS*) data as in k-space interpolation-based reconstruction methods such as GRAPPA, RAKI learns nonlinear convolutional neural networks from the *ACS* data. To interpolate the undersampled k-space, scan-specific *ACS* data is used to train the neural network. RAKI only aims to scan specifically, and does not require any training database of images. Compared with the early supervised learning algorithms, the training time of RAKI is reduced. Since RAKI is a single sampling method, Hosseini *et al.* [36] proposed self-consistent RAKI (sRAKI) that improves RAKI. It can realize arbitrary undersampling and is

This work was supported in part by National Natural Science Foundation of China under 61871206. (Corresponding authors: X. Xu, Q. Liu).

H. Tao, X. Xu and Q. Liu are with School of Information Engineering, Nanchang University, Nanchang 330031, China. (taohui@email.ncu.edu.cn, {xuxiaoling, liuqiegen}@ncu.edu.cn).

H. Wang, S. Wang and D. Liang are with Paul C. Lauterbur Research Center for Biomedical Imaging, SIAT, Chinese Academy of Sciences, Shenzhen 518055, China. ({hf.wang1, dong.liang}@siat.ac.cn, sophiasswang@hotmail.com).

an extension of the RAKI sampling method. Based on RAKI, Zhang et al. [37] proposed residual RAKI (rRAKI) which combines the advantages of linear convolution kernel in GRAPPA and nonlinear convolution kernel in RAKI. Later, Arefeen et al. [38] developed scan-specific artifact reduction in k-space (SPARK) which is a scan-specific model to reconstruct undersampled data, estimate and correct k-space errors. However, these methods have certain defects. For instance, the image reconstruction effect cannot reach the satisfactory state in the case of high acceleration rates, and they have the higher requirement on the amount of *ACS* lines.

Although the scan-specific learning exhibits great potentials in data flexibility for reconstruction, how to implicitly increase the amount of training data to leverage its performance is urgent. In this work, a novel multi-weight respecification of scan-specific learning is introduced. Concretely, by enforcing multi-weight respecification on the partially measured data, more data samples are available to guide the training procedure of scan-specific learning. This strategy leads to two possible benefits. Firstly, the multi-weight respecification explicitly widens the network structure, thus promoting the learning capability of the network. Secondly, by employing the multi-weight respecification with high-pass filters, it equals to explore different levels of high-frequency information from the measurement data to comprehensively alleviate the ill-poseness of the reconstruction task. This way is spiritually consistent to the compressed sensing (CS) [39]-[43] in the past two decades. Selecting the RAKI and rRAKI as the block bones, MW-RAKI and MW-rRAKI are formed by enforcing the strategy of multi-weight k-space respecification, respectively.

The main contributions of this study are as follows:
- The matrix weighting technique is employed in k-space measurement to leverage the effectiveness of scan-specific learning. It converts the object from k-space data to another feature data. By exploiting complementary information, the capability of representation information is improved.
- A multi-weight strategy is presented. More precisely, multi-weight training under different image support levels is aggregated for boosting the performance of our methods. Particularly, different implementations of image support levels may favor different image features, the multi-weight rule and average technique are jointly utilized to improve the robustness.
- In order to reduce the calculation amount and improve the performance efficiency, a more efficient network with the faster training speed is introduced to replace the naive RAKI network.

The rest of this article is organized as follows. Section II briefly introduces the related work of GRAPPA, RAKI, and rRAKI. Section III introduces the motivation of the paper, the influence of weight matrix on k-space and image domain, and the mathematical formulas of the present methods. Section IV validates the reconstruction performance of the proposed schemes, including experimental setup, reconstruction comparison and ablation research. Section V discusses parameters in MW-RAKI and MW-rRAKI, respectively. Section VI concludes the article.

## II. PRELIMINARIES

As stated in the introduction section, many deep learning techniques are performed in a manner that requires large amounts of MR samples. Therefore, it is necessary to develop a k-space reconstruction method that uses deep learning on a small amount of scan-specific *ACS* data.

GRAPPA is one of the most commonly used k-space interpolation methods for parallel imaging. For uniformly undersampled k-space data, GRAPPA uses linear convolutional kernels to estimate the missing data. Therefore, for the *i*-th coil k-space data $S_i$, we have:

$$S_i(k_x, k_y - m\Delta k_y) = \sum_{c=1}^{n_c} \sum_{b_x=-B_x}^{B_x} \sum_{b_y=-B_y}^{B_y} n_{i,m}(b_x, b_y, c) S_c(k_x - b_x \Delta k_x, k_y - R b_y \Delta k_y) \quad (1)$$

where $R$ is the acceleration rate, $S_i(k_x, k_y - m\Delta k_y)$ are the unacquired k-space lines, with $m = 1, \cdots, R-1$. $n_{i,m}(b_x, b_y, c)$ are the linear convolution kernels for estimating the data in the *i*-th coil specified by the location $m$ as above. $n_c$ is the number of coils. $S_c(k_x - b_x \Delta k_x, k_y - R b_y \Delta k_y)$ is a linear combination of acquired lines across all coils at the location $m$. $B_x$, $B_y$ are specified by the kernel size. The convolution kernels are estimated from scanning specific *ACS* data. Then, the missing data is obtained through the acquired points and the convolutional kernels.

Due to the linear convolution kernel, GRAPPA easily suffers from noise amplification at the high acceleration rate. In order to alleviate the deficiency, a nonlinear RAKI is proposed to estimate the missing k-space data by using the convolution neural network (CNN) with scan-specific [44], [45]. Its nonlinearity is introduced by the use of the rectified linear unit (ReLU) functions, and its linear part is simulated by the convolution layer. The training process of RAKI is similar to GRAPPA, which utilizes the *ACS* region to estimate the weight of the convolution kernel. $F_i$ is empowered by the series of functions $f_{i,m}$ in CNN. The relationship between the missing and acquired k-space data is as follows:

$$S_i(k_x, k_y - m\Delta k_y) = F_i(S_c(k_x - b_x \Delta k_x, k_y - R b_y \Delta k_y)) \quad (2)$$

The flowchart of the training process in RAKI is depicted in Fig. 1. The reference data of RAKI is obtained by moving the *ACS* region data of each coil. Actually, the training process of RAKI is a special self-supervised learning process. Its monitoring data comes from the conversion of the *ACS* region in the scanning coil. The training process needs to minimize the difference between the network output and reference data.

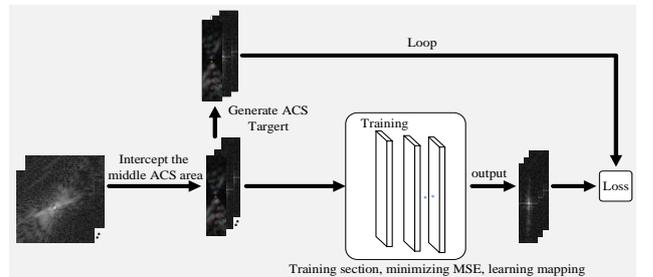

**Fig. 1.** The diagram flowchart of the training process in RAKI. The input of the network is all the sampled scan-specific *ACS* k-space data across all coils, while the referenced data is obtained by shifting the *ACS* lines.

After the network weights to be learned, uniform undersampled data is input into the network for reconstruction. The reconstructed channels are related to the acceleration factor. Fig. 2 visualizes the network architecture in RAKI for the process of k-space interpolation at the acceleration factor of 4. In this circumstance, the network will output the data of three channels. Because of the equal interval sampling, only one line is sampled every four lines, i.e., the first line is sampled, and the other three lines are missing lines. These missing lines are interpolated by the output of the network. One of the coils is reconstructed each time, and all the coils are reconstructed in turn.

Although RAKI improves noise resilience over linear methods, the linear convolutions still provide a sufficient baseline image quality and interpretability. In order to further improve the reconstruction quality, Zhang *et al.* utilized the residual network architecture to combine the advantages of linear and nonlinear RAKI reconstructions, called residual RAKI (rRAKI). It combines GRAPPA with RAKI in an interpretable manner, which not only retains the advantages of GRAPPA, but also maintains the scan-specificity of RAKI. The linear skip connection realizes linear interpolation that is similar to GRAPPA, while the multilayer nonlinear CNN component removes the noise amplification arising from the linear component. The training and reconstruction processes in rRAKI are similar to RAKI, and the differences are only reflected in the network structure.

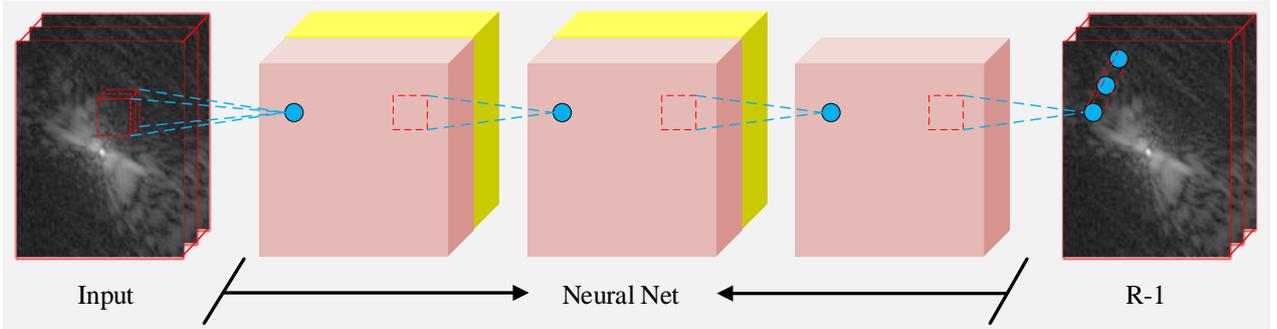

**Fig. 2.** Visualization of the network architecture in RAKI. Convolutional neural network (CNN) and activation function are represented pink and yellow, respectively. The input of the network is all the sampled k-space data across all coils, while the output is all missing lines for a given coil, leading to R-1 output channels.

## III. METHOD

### A. Motivation

From the theoretical viewpoint of machine learning, the amount of training data is heavily related to the number of constraints [46]. In the case of supervised end-to-end learning, more training data are required to match the increased network parameters that need to be determined [47], [48].

In the scan-specific learning method RAKI, only the scan-specific *ACS* are chosen as the training data. Since the amount of available *ACS* lines is limited, only deepening the network architecture, such as to augment the amount of network parameters, may not enhance the learning capability.

Instead, widening the network is an alternative to leverage the efficiency of scan-specific learning. A diagram of this strategy is shown in Fig. 3. On the one hand, widening the network will lead to more amounts of constraints in the learning formulation. It implicitly provides more training samples. On the other hand, the wider the network, the more the amount of network parameters are. A key question is that how do we widen the network. In this study, we achieve this goal via employing the multi-weight principle to k-space respecification.

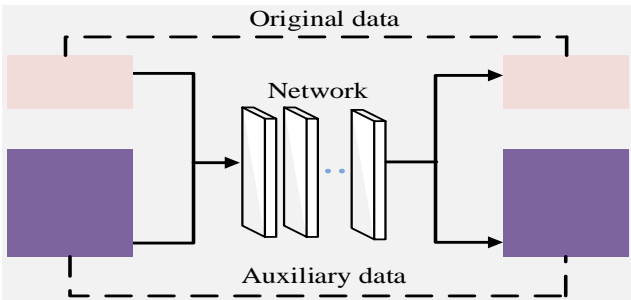

**Fig. 3.** Visual illustration of using the auxiliary data in scan-specific learning. Exploiting auxiliary data from the partially measured data not only provides more data samples, but also widens the network parameter space.

### B. Multi-weight Principle for K-space Respecification

As mentioned above, widening the network is an alternative to leverage the efficiency of scan-specific learning. If only the simple variables augmentation technique is used, the learning ability of the network is limited. Motivated by the works in [49], weighted matrices are designed in this study. Different implementations of weighted matrices may favor different characteristic information. Therefore, the multi-weight matrices are utilized to widen the network.

Specifically, multi-weight principle for k-space respecification leads to the following two potential benefits: First, as stated in compressed sensing community, image supports can be efficiently reduced by suppressing low-frequency information with the weighted technique. An image with small image support means fewer unknowns in unfolding equations (e.g., in the SENSE reconstruction). Second, the k-space data are multiplied by a weighted matrix to change the value range and extract appropriate image characteristics. As the corresponding output is the high-frequency feature after passing through the weighted matrices, the amplitude values of pixel locations in weighted k-space domain are much nearer and more homogeneous.

In order to better describe the weighted matrices, the formula is mathematically expressed as:

$$h(k_x, k_y) = M(k_x^2 + k_y^2)^P / D_0 \qquad (3)$$

where $k_x$ and $k_y$ are the pixel locations in two-dimensional region. $M$ and $D_0$ denote the amplitude parameter and the cut-off frequency, $P$ determines the smoothness of the filter boundary. As depicted in Fig. 4, the weighted matrix is similar to the high-pass filter [50], [51]. Weighted matrices should be designed such that the energy of low-frequency components is significantly reduced, and filtered *ACS* lines still provide enough information for calibration of convolution kernels.

A visualization of employing the high-pass filter to fully undersampled k-space data is demonstrated in Fig. 4. From Fig. 4(a), it can be seen that the brightness of the image is dramatically reduced after passing through the high-pass filter. Fig. 4(b) depicts the direct inverse Fourier transform (IFFT) of the filtered k-space data. The k-space data are multiplied by a weighted matrix to change the value range, which can become more uniform, and their range is closer. Simultaneously, the introduction of weighted technology largely decreases the amplitude.

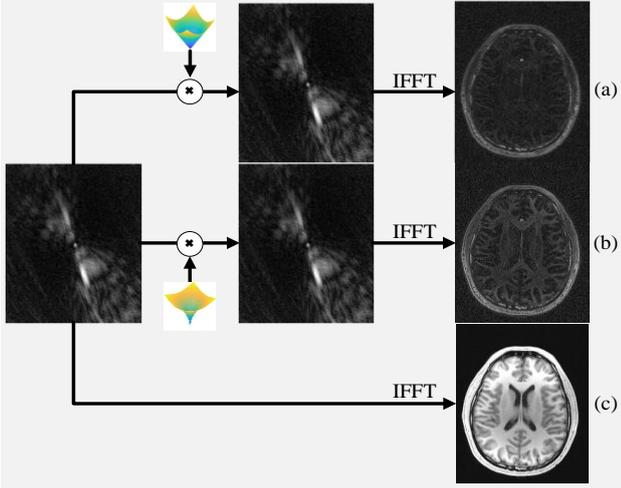

**Fig. 4.** Visualization of employing different high-pass weighting matrices on the k-space data. (a) Image with reduced image support by large high-pass filtering. (b) Image with reduced image support by small high-pass filtering. (c) The reference image.

## C. Proposed Methods

According to the multi-weight principle for k-space re-specification, we apply the multi-weight matrices strategy to the modified RAKI network and rRAKI network. Subsequently, MW-RAKI and MW-rRAKI are put forward, respectively.

*1) MW-RAKI:* Inspired by the relationship among data constraints, the network learning efficiency and network depth, we first modify the network architecture in the naïve RAKI to attain a more efficient network. Then, by enforcing the multi-weight learning strategy to the modified network, the final MW-RAKI based has been put forward. Same as the naïve RAKI, the convolutional kernels of MW-RAKI are determined by scan-specific sampled data. Meanwhile, MW-RAKI utilizes fewer network layers than RAKI, consisting of two layers of linear convolution and an activation function. A visualization of the training and testing process of MW-RAKI is depicted in Fig. 5.

More importantly, high-dimensional guiding strategy of multiple weighted matrices [52] in MW-RAKI involves two characteristics: (i) Learning some information in higher-dimensional space, rather than the original space. (ii) Using different implementations of image support levels. By incorporating the mathematical formulation (3) into the reconstruction equation (2), it yields:

$$S_i(k_x, k_y - m\Delta k_y) = F_i(\sum_{j=1}^{L} h_j \circ S_c(k_x - b_x\Delta k_x, k_y - Rb_y\Delta k_y)) \quad (4)$$

where $h_j$ represents the high-pass filter. It is worth noting that the all-pass filter $h_j = I$ is a special case of the high-pass filter.

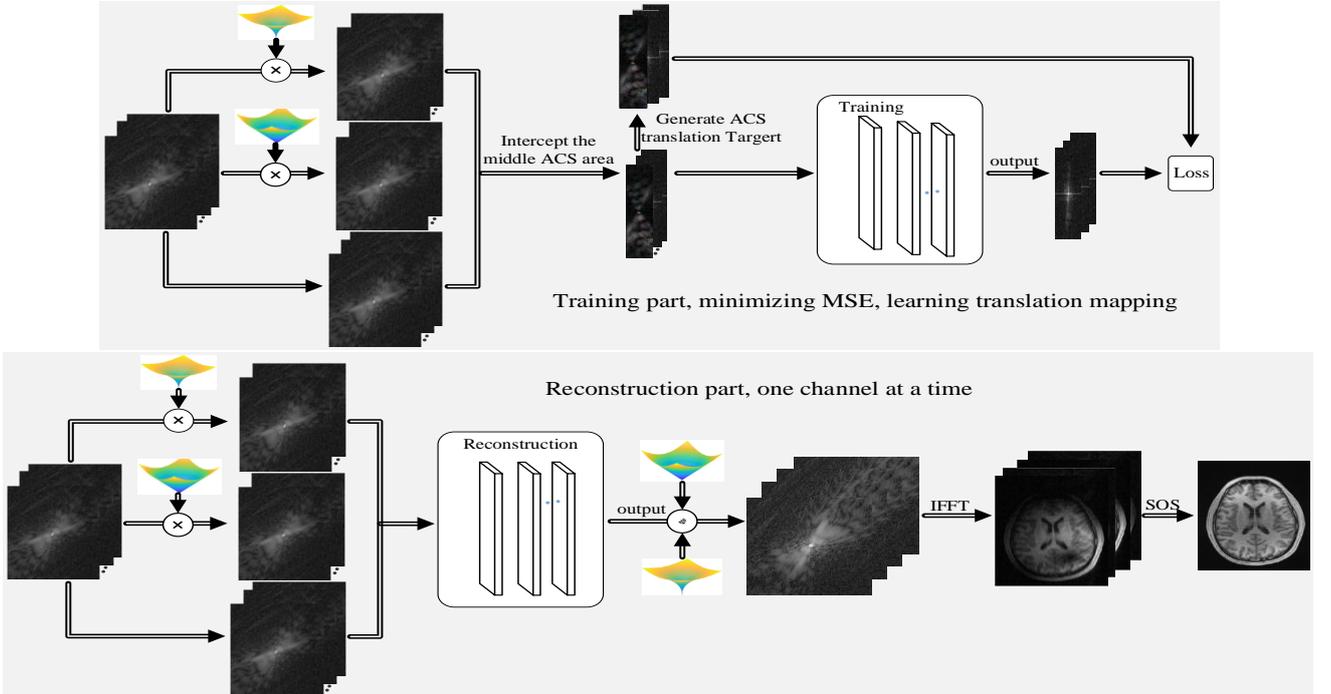

**Fig. 5.** The training and reconstruction flowchart of the proposed MW-RAKI. Top: The two filters and the auxiliary variable network scheme at the training stage. Bottom: This technique used for k-space interpolation scan-specific images at the restoration phase.

By utilizing the mean squared error (MSE), the training loss of MW-RAKI for channel $i$ network is given by:

$$L(\theta_i) = \| y_i - F_i(Y_{source}; \theta_i) \|_2^2 \quad (5)$$

where $\| \bullet \|_2$ denotes the $l_2$ norm. Let $y_i$ indicate the target points in $ACS$ region and the target is the missing point. $Y_{source}$ stands for the source points in the $ACS$ region. The source point is the point in the same position in all coils, which generally locates around the target point. The unknown network parameters, $\theta_i = \{w_1, w_2\}$ for that input channel need to be estimated. Similar to GRPPA, the pro-posed method uses the property of linear shift-invariant. It utilizes mathematical formulation (5) to learn convolutions in *ACS* region. Then, the convolutions are implemented to reconstruct the missing data by using Eq. (4). To update the convolutional kernels at iteration $t$, the parameters in the two-layer network are calculated as follows:

$$\overline{w}_i^{(t)} = \mu \overline{w}_i^{(t-1)} + \eta \partial L / \partial w_i^{(t-1)} \quad (6)$$

$$w_i^{(t)} = w_i^{(t-1)} - \overline{w}_i^{(t)} \quad (7)$$

where $\mu$ denotes the momentum rate, $\eta$ indicates the learning rate, $i \in \{1,2\}$, and backpropagation is used to calculate the derivative $\partial L / \partial w_i^{(t-1)}$. Simple gradient descent is used because of the limited size of the *ACS* region.

For the convenience of implementing the multi-weight principle for k-space respecification in MW-RAKI, some pre-process and post-process steps are needed: In the training process, we use the undersampled k-space data multiplied by the different weighted coefficient matrices and copy the original data as the input of the network. Particularly, the reference data is generated by translating the *ACS* data of each coil. In the reconstruction process, the input data are similar to the training process. The output of the network includes two parts. They are transformed to be complex-valued data and then are divided by the weighted matrix. Since all experimental data are multi-coil, the final consequence is obtained by merging these coil reconstructed images with sum-of-the-squares (SOS) operator.

*2) MW-rRAKI:* In addition to apply the multi-weight filtering strategy on RAKI, we further apply it to rRAK and form MW-rRAKI. The training and testing procedures of MW-rRAKI are similar to MW-RAKI, while the difference only occurs in the network structure. The proposed MW-rRAKI consists of a skip connection incorporating a linear convolution in parallel with three-layer CNNs.

More rigorously, the relationship between the estimated data and the acquired data in MW-rRAKI is formulated as follows:

$$S_i(k_x, k_y - m\Delta k_y) = F_i^*(\sum_{j=1}^{L} h_j \circ S_c(k_x - b_x \Delta k_x, k_y - Rb_y \Delta k_y)) + G_i(\sum_{j=1}^{L} h_j \circ S_c(k_x - b_x \Delta k_x, k_y - Rb_y \Delta k_y)) \quad (8)$$

Here only few variables in Eq. (8) are different from Eq. (4). Particularly, the network components $F_i^*$ and $G_i$ are trained jointly. $F_i^*$ are convolutional neural networks used by the original RAKI. $G_i$ represents the skip connection layer.

Finally, the training loss of MW-rRAKI for channel $i$ network is given by:

$$L(\theta_i) = \| y_i - F_i^*(Y_{source}; \theta_i) - G_i(Y_{source}; \theta_i) \|_2^2 \quad (9)$$

IV. EXPERIMENTS

A. Experiment Setup

*1) Datasets:* The experiments are performed on in-vivo 2D brain data from the MPRAGE dataset in [38] and SIAT dataset in [32].

In [38], the volunteers are scanned with a Siemens 3T Magnetom Skyra (Siemens Healthineers, Erlangen, Germany) system using a 32-channel receive head coil. More precisely, the first fully sampled MPRAGE scan is acquired with the following imaging parameters: FOV= $234 \times 188 \times 192$ $mm^3$, voxel size= $1 \times 1 \times 1$ $mm^3$, TR/TE/TI=2530/1.7/1100 $ms$, flip angle= $7°$, bandwidth=651 Hz/pixel. The raw data are Fourier transformed along the slice (third) dimension, and a single axial slice is used for subsequent experiments ($234 \times 188$ matrix size). The second MPRAGE dataset is fully phase encoded and oversampled by a factor of 3x along the readout dimension to accommodate wave-encoding. The imaging parameters are as follows: FOV=$256 \times 256 \times 192$ $mm^3$, voxel size=$1 \times 1 \times 1$ $mm^3$, TR/TE/TE=2500/3.48/1100 $ms$, flip angle=$8°$, readout duration is $5.08$ $ms$, maximum slew rate of wave gradients is 175 $mT/m/s$, maximum wave gradient amplitude is 9.4 $mT/m$, 15 sinusoidal wave cycles, k-space dimensions: $768 \times 256 \times 256 \times 32$. The wave point-spread function is calibrated with an auto-calibrated approach. Experiments are conducted on 2D wave-encoded data for a particular slice from the 3D dataset.

Besides, the brain data are scanned from a 3T Siemens's MAGNETOM Trio scanner using the T2-weighted turbo spin echo sequence [32]. The relevant imaging parameters are as follows: FOV= $220 \times 220$ $mm^2$, in-plane resolution= $0.9 \times 0.9$ $mm^2$, TR/TE=2500/149 $ms$, slice thickness= $0.86$ $mm$. Moreover, the number of coils is 12 and the collected dataset includes 500 MR data. In the experiments, only five complex-valued MR data with acquisition matrix of $256 \times 256$ are used to verify the performance of MW-RAKI and MW-rRAKI.

*2) Model Setting:* During the learning phase, we use undersampled MR images as the network input and handle it simultaneously via different weighted matrices. Adam is selected as an optimizer with the learning rate 0.001. To operate on complex-valued k-space data, we concatenate the real and imaginary portions of k-space along the so-called channel dimension of CNN inputs and outputs. Two convolutional kernels are trained for each output channel. At the training phase, CNNs are trained to interpolate missing k-space lines in multi-coil acquisitions using just filtered *ACS* data from the scan-specific for training. During the reconstruction phase, we use uniformly undersampled k-space data.

*3) Comparison Methods:* In the experiments, GRAPPA, RAKI, and rRAKI algorithms are compared with the proposed methods on three datasets. Two experimental scenarios are designed to verify the performance of the MW-RAKI and MW-rRAKI: The accelerated factors are varied over five values with *R*={4, 6, 8, 9, 10} and the *ACS* lines are varied over values with *ACS*={20, 24, 30, 36, 40}. For convenient reproducibility, the source code is available at: https://github.com/yqx7150/MW-RAKI-rRAKI.

*4) Implementation Details:* The sizes of the convolution kernels in MW-RAKI are $5 \times 2$ and $3 \times 2$. At the meantime, the sizes of the convolution kernels in MW-rRAKI are $5 \times 2$, $1 \times 1$ and $3 \times 2$ sequentially. The learning parameters of RAKI, rRAKI, MW-RAKI, and MW-rRAKI networks are the same, and the kernel size of GRAPPA is $3 \times 3$. The configuration of the server based on this experiment is as follows: 2 NVIDIA Titan XP GPUs, 16GB RAM, 1080Ti, the system of ubuntu 16. RAKI, rRAKI, MW-RAKI and MW-rRAKI are implemented by Python 3.7 and TensorFlow 2.0.0, supported by CUDA 10.0 and CuDNN 7.6.4. Python environment is created by Anaconda 3.8.3. GRAPPA is real-ized by using Matlab R2017b.

*5) Evaluation Metrics:* To quantitatively evaluate the quality of reconstructed images, three metrics are utilized: Peak signal to noise ratio (PSNR), structural similarity (SSIM) and root mean square error (RMSE).

### B. Results on 2D MPRAGE Scan

To validate the performance of MW-RAKI and MW-rRAKI, undersampled data with $R$=6 and $ACS$=40 is reconstructed in the experiment. Fig. 6 visualizes the reconstruction results of GRAPPA, RAKI, rRAKI, MW-RAKI, and MW-rRAKI with enlarged views and error maps. Particularly, the error diagram is magnified 5 times. As illustrated in Fig. 6, the visual quality of reconstruction by different methods is different. The proposed methods can achieve satisfactory results with clearer contours, sharper edges, and finer image details. From the error maps, we can observe that both MW-RAKI and MW-rRAKI can preserve the subtle structures of images. In the enlarged image, the image quality of MW-RAKI and MW-rRAKI is clearer.

Table I summarizes the quantitative PSNR, SSIM and RMSE results of different methods with $R$={4, 6, 8} and $ACS$={36, 40}. Intuitively, the average PSNR value of the results obtained by MW-RAKI is nearly 2 dB higher than that in RAKI. Taking the case of $R$=4 for instance, the highest PSNR values achieved by MW-RAKI and MW-rRAKI are 37.12 dB and 37.21 dB, which are higher than the values of 35.47 dB, 35.80 dB and 27.39 dB obtained by GRAPPA, RAKI and rRAKI. It can be observed that MW-RAKI and MW-rRAKI yield higher values at the majority of the acceleration factors. The superiority of the proposed methods over the competing algorithms is more striking under higher acceleration rates such as $R$=8. In this circumstance, RAKI, rRAKI, MW-RAKI, and MW-rRAKI have the higher signal-to-noise ratio than GRAPPA. Furthermore, when $ACS$ is decreased to be 36, the quantitative indexes of MW-RAKI and MW-rRAKI are still the best.

Both quantitative and qualitative assessments demonstrate that the proposed methods offer significant improvements in reconstruction quality at highly undersampling rates. The trained network can use diverse image support level signal to retain different detailed structure and texture.

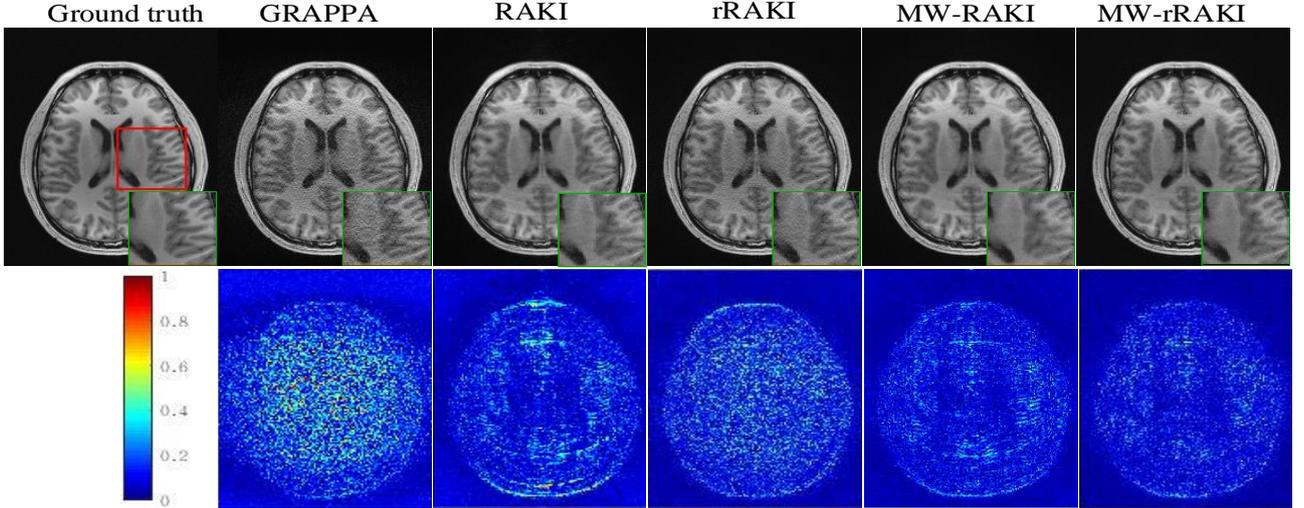

**Fig. 6.** Reconstruction comparison on uniform sampling at acceleration factor $R$=6 and $ACS$=40 on 2D MPRAGE. Top: Reference, reconstruction by GRAPPA, RAKI, rRAKI, MW-RAKI, and MW-rRAKI. Bottom: The error maps that are magnified by 5 times. Green boxes illustrate the zoom in results.

TABLE I
RECONSTRUCTION PSNR, SSIM AND RMSE VALUES OF SEVERAL METHODS ON MPRAGE UNDER UNIFORM UNDERSAMPLING WITH DIFFERENT SAMPLING RATES AND *ACS* LINES. ALL THE RESULTS ARE AVERAGED VALUES BY 4 IMPLEMENTATIONS.

| MPRAGE, $ACS$=40 | GRAPPA | RAKI | rRAKI | MW-RAKI | MW-rRAKI |
|---|---|---|---|---|---|
| $R$=4 | 35.47/0.9618/4.51 | 35.80/0.9698/4.33 | 36.71/0.9779/3.90 | 37.12/0.9830/3.72 | **37.21/0.9836/3.69** |
| $R$=6 | 27.61/0.8557/11.13 | 32.50/0.9499/6.27 | 31.18/0.9400/7.79 | 33.50/0.9655/5.64 | **33.66/0.9660/5.54** |
| $R$=8 | 19.00/0.6294/30.00 | 27.39/0.9016/11.43 | 25.67/0.8695/13.92 | 29.99/**0.9382**/8.46 | **30.06**/0.9360/**8.42** |
| MPRAGE, $ACS$=36 | GRAPPA | RAKI | rRAKI | MW-RAKI | MW-rRAKI |
| $R$=4 | 35.39/0.9610/4.55 | 35.75/0.9689/4.36 | 36.58/0.9787/3.96 | 36.50/0.9822/4.00 | **36.95/0.9824/3.80** |
| $R$=6 | 27.40/0.8610/11.41 | 31.25/0.9435/7.32 | 30.68/0.9350/7.82 | 32.46/**0.9580**/6.38 | **32.58**/0.9575/**6.28** |
| $R$=8 | 18.43/0.6142/32.02 | 26.36/0.8849/12.86 | 25.42/0.8555/14.33 | 29.02/0.9266/9.47 | **29.32/0.9253/9.15** |

### C. Results on Wave-encoded MPRAGE

Wave-encoded MPRAGE dataset is adopted to further verify the performance. The slices are retrospectively undersampled at $R$={4, 6, 8} with $ACS$={20, 24, 30, 36, 40}. It is well known that the GRAPPA reconstruction involves more noise. RAKI and rRAKI contain more details than GRAPPA. The reconstructions as well as zoom-in results attained by different algorithms are presented in Figs.7-8. The error maps are illustrated through the color bar. In the enlarged region, our methods provide more realistic quality and retain the detailed structure as well as texture. Additionally, the proposed methods have lower error maps.

Table II lists the average quantitative results of reconstructed images under different sampling rates in wave-encoded MPRAGE. As can be observed, MW-RAKI and MW-rRAKI outperform GRAPPA, RAKI, and rRAKI significantly. The average PSNR results obtained by MW-RAKI and MW-rRAKI is about 1 dB higher than those of RAKI. Theoretically, the inferiority of GRAPPA indicates that the traditional reconstruction methods are insufficient to tackle with high acceleration rate. When the acceleration rate increases from 4 to 8, the advantages of the proposed methods have become even more pronounced. Although RAKI and rRAKI based on deep learning methodology are much better than GRAPPA, they still have large rooms to improve. In addition, the PSNR values of the proposed algorithms at *ACS*=36 are better than those of RAKI at *ACS*=40. Compared with the competing algorithms, MW-RAKI and MW-rRAKI produce the highest SSIM, PSNR and lowest RMSE values at *ACS*=36 and *R*=8.

Fig. 9 compares RMSE values of RAKI, rRAKI, MW-RAKI, and MW-rRAKI results at *R*= 6 and varying *ACS* sizes. At all acceleration rates and *ACS* sizes, MW-RAKI and MW-rRAKI perform better than RAKI and rRAKI. Particularly, MW-RAKI and MW-rRAKI achieve improved robustness to *ACS* sizes lower than 25 lines.

*D. Results on SIAT Data*

To investigate the performances of various methods at high acceleration rates, comparison experiments are implemented on 5 complex-valued k-space data from SIAT dataset. GRAPPA has visible residual artifacts at high acceleration rate, thus is not compared here.

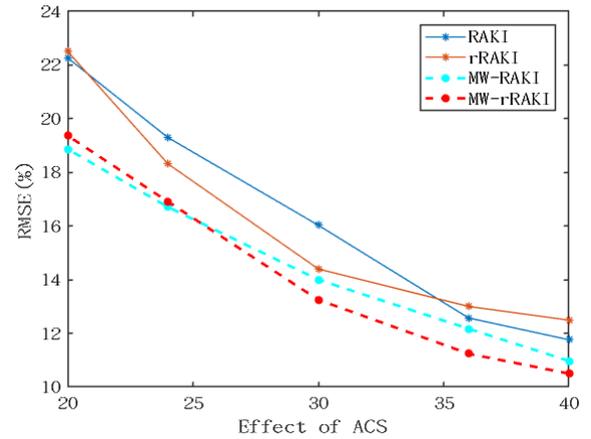

**Fig. 9.** Comparisons between GRAPPA, RAKI, rRAKI, MW-RAKI and MW-rRAKI for *R*=6 and a range of *ACS* sizes. MW-RAKI and MW-rRAKI always perform at least as well or better than RAKI, particularly at smaller *ACS* sizes.

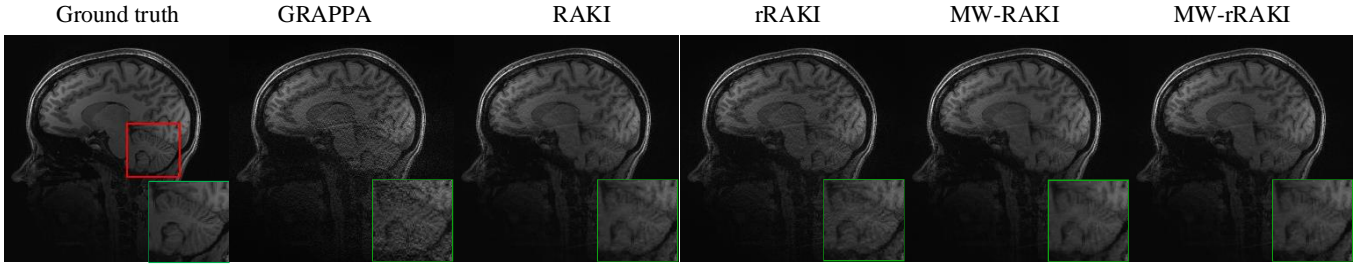

**Fig. 7.** Visual comparisons under uniform sampling at acceleration factor *R*=6 and *ACS*=40 on 2D MPRAGE. From left to right: Reference, reconstruction of GRAPPA, RAKI, rRAKI, MW-RAKI, and MW-rRAKI. Green boxes illustrate the zoom in results.

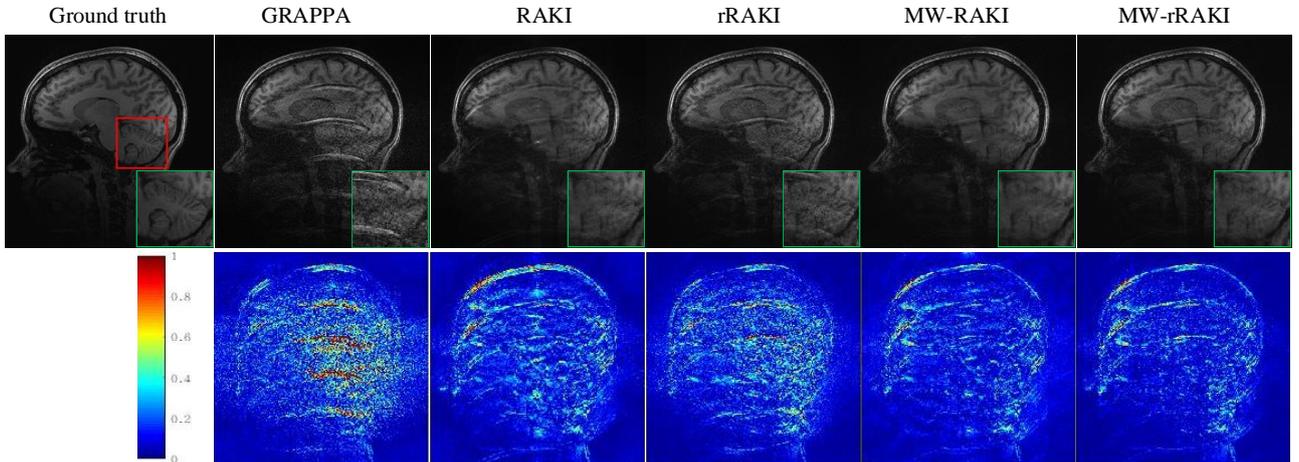

**Fig. 8.** Performance comparison of different methods at acceleration factor *R*=6 and *ACS*=20. Top: Reference, reconstruction by GRAPPA, RAKI, rRAKI, MW-RAKI, and MW-rRAKI. Bottom: The error maps that are magnified by 5 times. Green boxes illustrate the zoom in results.

TABLE II
COMPARISON ON RESULTS OF SEVERAL METHODS. UNDER UNIFORM UNDERSAMPLING, DIFFERENT SAMPLING RATES, *ACS*=40 AND *ACS*=36, DIFFERENT ALGORITHMS ARE USED TO RECONSTRUCT THE AVERAGE PSNR, SSIM AND RMSE VALUES ON WAVE-ENCODED MPRAGE. ALL THE RESULTS ARE AVERAGED VALUES BY 4 IMPLEMENTATIONS.

| Wave-encoded MPRAGE, *ACS*=40 | GRAPPA | RAKI | rRAKI | MW-RAKI | MW-rRAKI |
|---|---|---|---|---|---|
| *R*=4 | 37.19/0.9587/9.25 | 37.65/0.9646/8.70 | 38.08/0.9709/8.35 | **39.26/0.9804/7.28** | 39.19/**0.9824**/7.35 |
| *R*=6 | 30.95/0.8811/18.97 | 35.12/0.9448/11.73 | 34.58/0.9433/12.48 | 35.81/0.9615/10.84 | **36.07/0.9625/10.51** |
| *R*=8 | 26.49/0.7906/31.07 | 30.90/0.9058/19.07 | 30.94/0.8963/18.97 | **32.63/0.9376/15.63** | 32.58/0.9346/15.72 |
| Wave-encoded MPRAGE, *ACS*=36 | GRAPPA | RAKI | rRAKI | MW-RAKI | MW-rRAKI |
| *R*=4 | 37.04/0.9572/9.40 | 37.57/0.9644 /8.85 | 37.99/0.9713/8.43 | **39.19/0.9796/7.34** | 39.13/**0.9822**/7.4 |
| *R*=6 | 30.20/0.8685/20.67 | 34.52/0.9414/12.57 | 34.24/0.9397/12.98 | 34.79/0.9581/12.19 | **35.53/0.9588/11.2** |
| *R*=8 | 25.23/0.7621/36.63 | 29.50/0.8871/22.4 | 29.83/0.8793/21.58 | 31.60/**0.9272**/17.60 | **31.60**/0.9228/**17.59** |

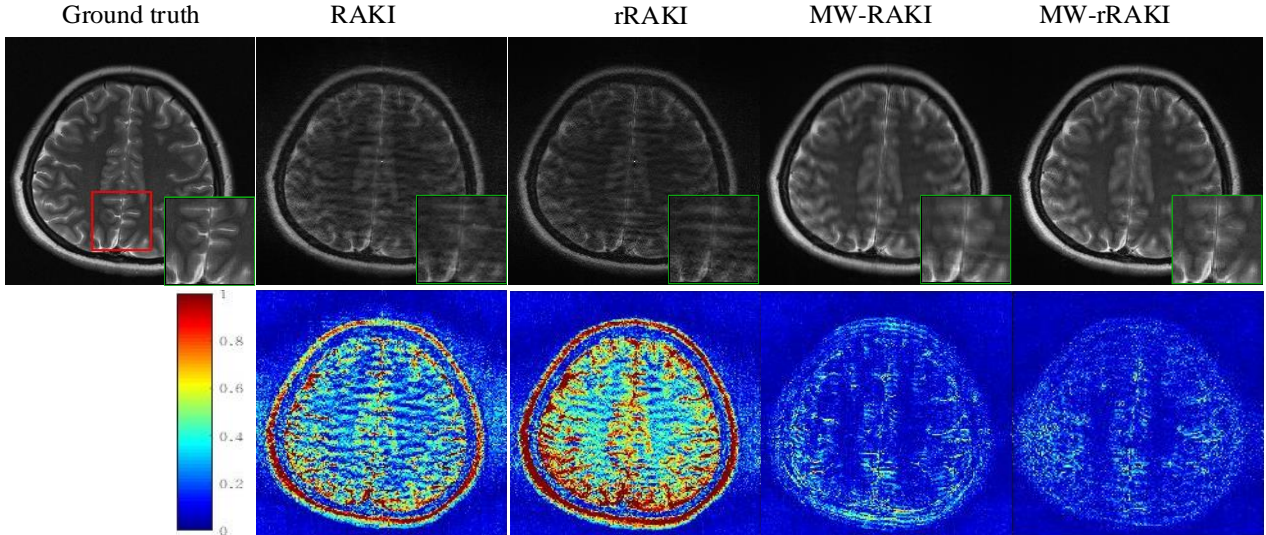

**Fig. 10.** Reconstructed results using different methods at acceleration factor *R*=9 and *ACS*=40. Top: Reference, reconstruction by RAKI, rRAKI, MW-RAKI, and MW-rRAKI. Bottom: The error maps that are magnified by 5 times. Green boxes illustrate the zoom in results.

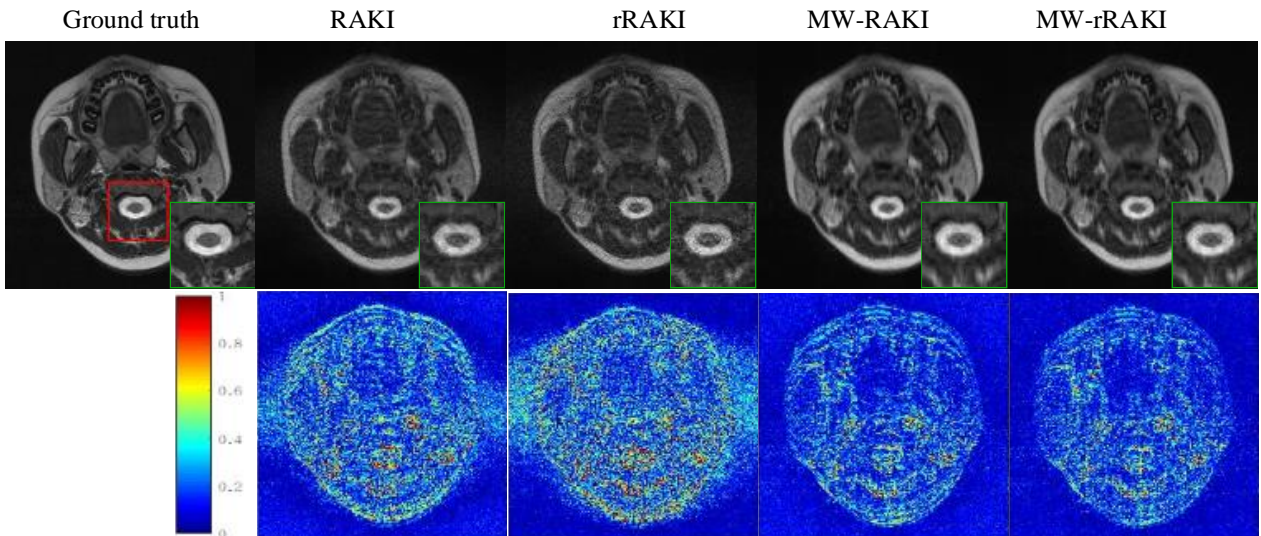

**Fig. 11.** Reconstruction comparison on uniform sampling at acceleration factor *R*=10 and *ACS*=40. Top: Reference, reconstruction by RAKI, rRAKI, MW-RAKI, and MW-rRAKI. Bottom: The error maps that are magnified by 5 times. Green boxes illustrate the zoom in results.

Visualization results reconstructed by different methods on uniform undersampling under acceleration factors $R$=9 and $R$=10 are provided in Figs. 10-11. It is easy to see that our methods yield better reconstruction results. Due to the linear component in rRAKI, it is more vulnerable to noise at high acceleration rates. Visually, rRAKI has the worst error map. In conclusion, the proposed methods retain more details and are faithfully closer to the reference image.

Quantitative results of different methods under high acceleration rates are listed in Table III. As can be observed, the PSNR values of MW-RAKI and MW-rRAKI are at least 5dB higher than those of RAKI and rRAKI. For instance, when the acceleration factor is 9, the PSNR value of MW-rRAKI is 29.37 dB, while the PSNR values of RAKI and rRAKI are 24.00 dB and 22.79 dB, respectively.

### E. Ablation Study

Table IV details the quantitative results of the number of different weighting filters. When the number of high-pass filters increases, the PSNR and SSIM values of the proposed methods are promoted accordingly. When the number increases to about 2, the improvement growth of slows down.

Considering that the reconstruction time will also gradually increase as the number of high-pass filters becomes larger, we use two filter combinations in almost experiments.

TABLE III
AVERAGE PSNR, SSIM AND RMSE VALUES OF RECONSTRUCTING 5 TEST DATA BY DIFFERENT ALGORITHMS AT HIGH ACCELERATION RATES AND UNIFORM UNDERSAMPLING WITH *ACS*=40 IN SIAT.

| R | 9 | | | 10 | | |
|---|---|---|---|---|---|---|
| Method | PSNR | SSIM | RMSE | PSNR | SSIM | RMSE |
| RAKI | 24.00 | 0.8118 | 26.73 | 23.06 | 0.7972 | 29.46 |
| rRAKI | 22.79 | 0.7888 | 30.48 | 21.73 | 0.7659 | 34.55 |
| MW-RAKI | 29.20 | 0.8967 | 14.33 | 29.04 | 0.8965 | 14.59 |
| MW-rRAKI | **29.37** | **0.8993** | **14.02** | **29.22** | **0.8979** | **14.25** |

TABLE IV
INFLUENCE OF THE WEIGHTING FILTER NUMBER ON RECONSTRUCTION RESULTS UNDER UNIFORM SAMPLING WITH *ACS*=40, *P*=0.4 AND *R*=6.

| Filter number | 1 | 2 | 3 | 4 |
|---|---|---|---|---|
| *R*=4 | 36.63 | 37.15 | **37.19** | 37.03 |
|  | 0.9816 | 0.9829 | **0.9833** | 0.9830 |
|  | 3.94 | **3.17** | 3.70 | 3.72 |
| *R*=6 | 33.27 | 33.54 | 33.60 | **33.65** |
|  | 0.9640 | 0.9656 | **0.9664** | **0.9664** |
|  | 5.81 | 5.63 | 5.58 | **5.59** |

TABLE V
INFLUENCE OF P ON RECONSTRUCTION RESULTS UNDER ACS=40.

| P | 0.2 | 0.3 | 0.4 | 0.5 | 0.6 |
|---|---|---|---|---|---|
| R=4 | 35.88/0.9706/4.30 | 36.00/0.9717/4.24 | 36.19/0.9738/4.15 | 36.19/0.9769/4.15 | **36.24/0.9790/4.13** |
| R=6 | 32.76/0.9550/6.15 | 32.73/0.9545/6.17 | **32.83**/0.9552/**6.11** | 32.63/0.9578/6.25 | 32.75/**0.9594**/6.16 |

TABLE VI
INFLUENCE OF DIFFERENT COMBINATIONS OF P ON RECONSTRUCTION RESULTS UNDER ACS=40.

| $(P_1, P_2)$ | (0.6, 0.5) | (0.6, 0.4) | (0.6, 0.3) | (0.6, 0.2) | (0.5, 0.4) |
|---|---|---|---|---|---|
| R=4 | 36.76/0.9832/3.89 | 37.12/0.9831/3.72 | 37.01/0.9832/3.77 | **37.24/0.9835/3.67** | 37.07/0.9832/3.75 |
| R=6 | 33.40/**0.9664**/5.71 | **33.52**/0.9660/**5.64** | 33.37/0.9660/5.74 | 33.42/0.9662/5.70 | 33.46/0.9661/5.68 |

TABLE VII
INFLUENCE OF THE NUMBER OF NETWORK LAYERS ON THE RECONSTRUCTION RESULTS UNDER R=6.

| Layer number | RAKI, 5 | RAKI, 3 | RAKI, 2 | RAKI, 1 | MW-RAKI, 5 | MW-RAKI, 3 | MW-RAKI, 2 | MW-RAKI, 1 |
|---|---|---|---|---|---|---|---|---|
| ACS=40 | 32.38 / 0.9491 / 6.43 | 32.50 / 0.9491 / 6.34 | 32.54 / 0.9528 / 6.31 | 30.09 / 0.9247 / 8.37 | 32.00 / 0.9560 / 6.72 | 32.97 / 0.9576 / 6.01 | **33.54 / 0.9656 / 5.63** | 32.66 / 0.9458 / 6.22 |
| ACS=36 | 31.09 / 0.9423 / 7.46 | 31.31 / 0.9438 / 7.28 | 31.52 / 0.9462 / 7.10 | 29.80 / 0.9206 / 8.65 | 29.92 / 0.9435 / 8.50 | 31.93 / 0.9508 / 6.77 | **32.48 / 0.9579 / 6.36** | 32.22 / 0.9420 / 6.55 |

Table V presents the quantitative results of different filter coefficients *P* under *ACS*=40. Different filter coefficients have different shapes of corresponding filters. The filter coefficient is larger, the corresponding low-frequency components pass through less. Additionally, quantitative results of different combinations on filter coefficients are listed in Table VI. As can be seen, the results of the multiple combinations in Table VI are better than the results of the single filter in Table V.

Table VII lists the quantitative results of different network depths. It can be observed that the results of the three-layer CNNs and the two-layer CNNs have little difference in RAKI. While the two-layer CNNs outperforms much that of the three-layer CNNs in MW-RAKI. Therefore, the two-layer CNNs are selected for reconstruction in this work.

Fig. 12 depicts the iterative tendency of the reconstruction RMSE curve. Interestingly, the proposed MW-RAKI converges faster with the curve drops quickly at early iterations. It has already converged after about 250 iterations, while RAKI has a slow convergence rate until about 1000 iterations. The over-fitting phenomenon will appear when both of them continue training after convergence.

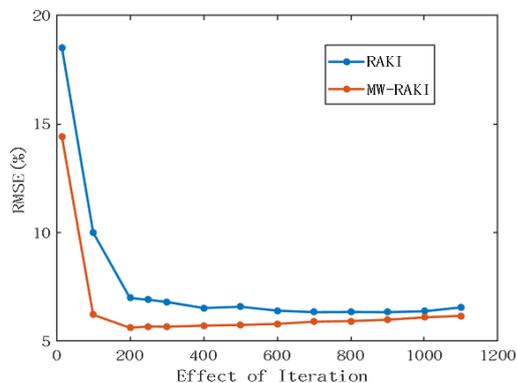

**Fig. 12.** Quantitative RMSE curve vs. iteration index.

## V. DISCUSSIONS

In this study, the underlying idea of applying weighted matrix provides better noise resilient property than GRAPPA, and fewer blurring artifacts than RAKI and rRAKI at highly undersampling rates. In this subsection, we discuss some possible ways to further leverage the performance of the proposed methods.

This work makes full use of the advantages of data with diversity. Therefore, we investigate different ways to concatenate the k-space data of the network input. we find that concatenating data along the first (batch) dimension achieves better the performance of reconstruction than other dimensions. How to design the optimal data structure as the network input for efficient self-supervised learning is necessary.

In the case of the high acceleration rates, RAKI and rRAKI need larger *ACS* region to better estimate the convolution kernels. In this study, we attempt to reduce the number of network layers and widen the network width to reduce the requirements of *ACS* lines. How to incorporate other strategies such as the virtual conjugate coil (VCC) [53] into the weighting matrix principle to further reduce the *ACS* lines is worth investigation in further study.

## VI. CONCLUSIONS

In this study, a new scan-specific model was proposed to accelerate parallel MRI by multi-weight matrices [53]-[56]. More specifically, multi-weight principle for k-space respecification provided more diverse and homogeneous representation of the data structure. Subsequently, imposing higher-dimensional structural information learned by network supported efficient scan-specific learning. Additionally, reducing the number of network layers avoided possible estimation inaccuracy due to insufficient of available data samples. Both qualitative and quantitative experimental results demonstrated that the proposed MW-RAKI and MW-rRAKI achieved better performances over the traditional linear reconstruction techniques and the recent nonlinear data-driven learning algorithms.